\documentclass{PoS}

\usepackage{axodraw}

\def\bfi{\begin{figure}}
\def\efi{\end{figure}}
\def\btab{\begin{table}}
\def\etab{\end{table}}
\def\bce{\begin{center}}
\def\ece{\end{center}}

\newcommand{\rT}{{\mathrm{T}}}

\def\ga{\gamma}
\def\de{\delta}

\def\si{\sigma}

\def\De{\Delta}

\def\reffi#1{\mbox{Figure~\ref{#1}}}

\def\refta#1{\mbox{Table~\ref{#1}}}

\def\citere#1{\mbox{Ref.~\cite{#1}}}
\def\citeres#1{\mbox{Refs.~\cite{#1}}}

\newcommand{\GeV}{\unskip\,\mathrm{GeV}}

\newcommand{\TeV}{\unskip\,\mathrm{TeV}}

\def\mathswitch#1{\relax\ifmmode#1\else$#1$\fi}
\def\mathswitchr#1{\relax\ifmmode{\mathrm{#1}}\else$\mathrm{#1}$\fi}
\def\mathswitchit#1{\relax\ifmmode{#1}\else$#1$\fi}

\newcommand{\PW}{\mathswitchr W}

\newcommand{\PH}{\mathswitchr H}

\newcommand{\Pe}{\mathswitchr e}

\newcommand{\Pd}{\mathswitchr d}

\newcommand{\Pu}{\mathswitchr u}
\newcommand{\Ps}{\mathswitchr s}
\newcommand{\Pc}{\mathswitchr c}
\newcommand{\Pb}{\mathswitchr b}

\newcommand{\Pp}{\mathswitchr p}

\newcommand{\Pep}{\mathswitchr {e^+}}
\newcommand{\Pem}{\mathswitchr {e^-}}

\newcommand{\PWp}{\mathswitchr {W^+}}
\newcommand{\PWm}{\mathswitchr {W^-}}

\newcommand{\MW}{\mathswitch {M_\PW}}

\def\ie{i.e.\ }
\def\eg{e.g.\ }

\newcommand{\mvev}{\nu_\mu\mu^+ \Pe^-\bar\nu_\Pe}
\newcommand{\mc}{\mathcal}

\newcommand{\ed}{\end{document}}

\newcommand{\mr}[1]{{\mathrm{#1}}}

\title{Next-to-leading-order electroweak corrections to pp$\to$WW$\to$~4~leptons in double-pole approximation at~the~LHC}

\ShortTitle{NLO EW corrections to pp$\to$WW$\to$~4~leptons in double-pole approximation at~the~LHC}

\author{\speaker{Marina Billoni}\\
        Johannes Gutenberg-Universit\"at Mainz\\
        E-mail: \email{billoni@uni-mainz.de}}

\author{Stefan Dittmaier\\
        Albert-Ludwigs-Universit\"at Freiburg\\
        E-mail: \email{stefan.dittmaier@physik.uni-freiburg.de}}

\author{Barbara J\"ager\\
        Johannes Gutenberg-Universit\"at Mainz\\
        E-mail: \email{jaeger@thep.physik.uni-mainz.de}}

\author{Christian Speckner\\
        Albert-Ludwigs-Universit\"at Freiburg\\
        E-mail: \email{christian.speckner@physik.uni-freiburg.de}}

\abstract{We report on our calculation of next-to-leading-order electroweak corrections to W-boson pair production at the LHC, taking into account off-shell effects and spin correlations of the W bosons and their leptonic decays in the framework of a double-pole approximation. We study the various contributions to the electroweak corrections in detail and discuss their impact on selected observables within a realistic event-selection setup.}

\FullConference{11th International Symposium on Radiative Corrections (Applications of Quantum Field Theory to Phenomenology) \\
		 22-27 September 2013\\
		 Lumley Castle Hotel, Durham, UK}

\begin{document}

\section{Introduction}
The comparison of measured $\PW$-boson pair production cross sections and related observables at the CERN Large Hadron Collider (LHC)
to precise theoretical predictions allows us to probe the non-abelian structure of the electroweak sector within the Standard Model (SM).
Due to the sensitivity of this process to triple weak-gauge-boson vertices,
exclusion limits on anomalous gauge couplings can be extracted \cite{ATLAS:2012mec,Chatrchyan:2013oev}. 
In the light of the Higgs decay channel $\PH\to\PW\PW^{\star}$, four-lepton production 
is an irreducible background whose impact on the Higgs signal has to be well understood.
In order to avoid that unknown higher-order corrections are misinterpreted as traces of ``new physics''
the calculation of these corrections has to be refined and extended whenever possible.

To this end, next-to-leading order (NLO) QCD corrections have been calculated
many years ago \cite{nloqcd}. On the QCD side various improvements beyond NLO exist (see \eg \citere{Campanario:2013wta} and references therein), and
currently a lot of effort is put into the calculation of next-to-next-to-leading order (NNLO) QCD corrections where so far only partial
results in the high-energy limit are known \cite{nnloqcd}. 

For the energy domain accessible at the LHC it is well known that electroweak (EW) corrections can become important as they are
enhanced by large logarithms at high scales. Recently, the EW corrections to on-shell production of weak-gauge boson pairs have been
calculated \cite{Bierweiler:2012kw,Baglio:2013toa} and found to be sizable.
Our calculation refines the existing theoretical predictions on EW corrections to $\PW$-boson pair production
as we consider $\mc{O}(\alpha)$ corrections to the four-lepton final state $\mvev$ in the so-called
{\em double pole approximation}~(DPA) following the {\tt RacoonWW} approach~\cite{racoon}, which was
developed to describe $\PW$-pair production in $\Pep\Pem$ annihilation at LEP2.

\section{Four-lepton production at the LHC}
Before going into a detailed discussion of the impact of individual contributions
to the NLO EW corrections this section gives a brief overview
of the various contributions included in our calculation. For technical issues we refer to 
\citere{Billoni:2013aba} where all aspects of the involved calculation are presented in detail.

\subsection{Leading-order contributions}
%
\bfi
\centerline{
\includegraphics[scale=0.8]{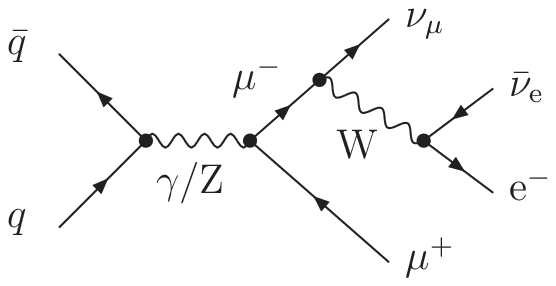}
\hspace{2em}
\includegraphics[scale=0.8]{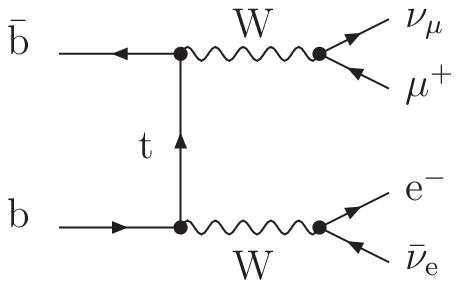}
\hspace{2em}
\includegraphics[scale=0.8]{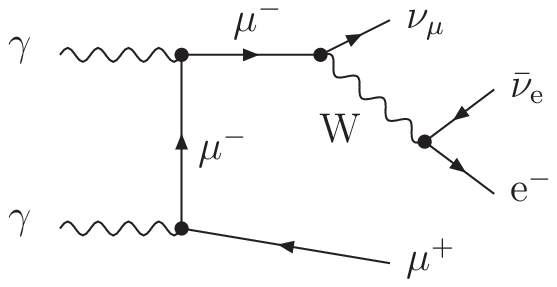}
}
\caption{Representative Feynman diagrams for the three different leading-order contributions:
$\bar qq$ channel (left), $\bar\Pb\Pb$ channel (middle), and  $\ga\ga$ channel (right).}
\label{fig:lo-graphs}
\efi
%
At leading order (LO) we encounter three different production channels leading to the final state $\mvev$ at the LHC.
The dominant production channel consists of antiquark--quark annihilation,
\bce
$\bar q q \to \mvev\,$,
\ece
with $q=\Pu,\Pd,\Pc,\Ps\,$. If the incoming quark pair consists of bottom quarks, a top quark appears as an intermediate state.
For this reason we separate this case from the previous one, 
\bce
$\bar \Pb \Pb \to \mvev\,$.
\ece
The third contribution consists of the photon--photon induced subprocess,
\bce
$\gamma\gamma \to \mvev\,$.
\ece
\reffi{fig:lo-graphs} shows one representative Feynman diagram for each contributing LO processes.
In all three cases our predictions at LO are based on full $2\to 4$ matrix elements, so that off-shell effects
are fully taken into account. As the contributions of the $\bar\Pb\Pb$- and $\ga\ga$-induced subprocesses
are expected to be small due to the small distribution functions of their incoming partons,
our calculation of EW corrections is restricted to the antiquark--quark
induced subprocess. For the discussion of our numerical results we always use the LO prediction of
the $\bar q q$-induced subprocess as normalization and all other contributions as well as EW corrections
are often presented in terms of relative corrections to this contribution. In this sense we introduce for the two 
additional tree-level contributions discussed here the relative corrections $\de_{\bar\Pb\Pb}$ and $\de_{\ga\ga}$,
respectively.

\subsection{NLO EW corrections}
%
\bfi
\centerline{
\includegraphics[scale=0.8]{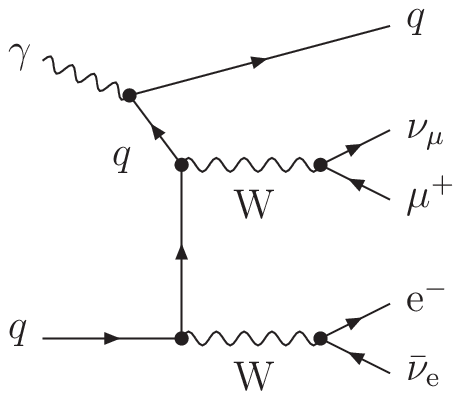}
\hspace{5em}
\includegraphics[scale=0.8]{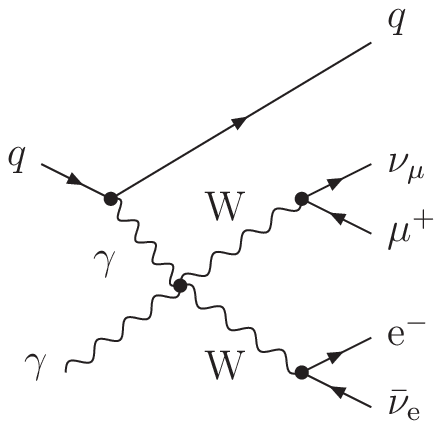}
}
\caption{Representative Feynman diagrams for the real-emission contribution $\ga q\to\mvev q$.}
\label{fig:real}
\efi
%
As mentioned above, the calculation of NLO EW corrections is relevant only for the antiquark--quark
induced subprocess. Regarding virtual electroweak corrections to $\bar q q \to \mvev$
a systematic expansion around the resonance
poles of the two $\PW$ bosons leads to so-called {\it factorizable} and {\it non-factorizable} corrections in DPA.
The calculation of these corrections is much simpler compared to the evaluation of the full electroweak corrections
and results in an efficient computer code that allows for fast numerical computation.
For $\Pep\Pem$ annihilation, results obtained in DPA have been compared to the full
calculation of EW corrections to $\Pep\Pem\to 4$ fermions \cite{Denner:2005es}. The corrections beyond DPA
are found to be below $0.5\%$ for moderate scattering energies and reach $1-2\%$ for a centre-of-mass energy in the TeV range. 
This result confirms the naive error estimate that the terms neglected in DPA are suppressed by a factor
$\alpha/\pi\times \Gamma_\PW/\MW$ with respect to the LO results, at least in energy domains where both gauge bosons
are near their mass shell.

However, this approximation is valid only for partonic centre-of-mass energies
sufficiently above the threshold for on-shell $\PW$-pair production, \ie for energies
larger than twice the $\PW$-boson mass. Thus for energies below this threshold we use a so-called 
{\em improved Born approximation}~(IBA), which is based on leading universal corrections.

Our evaluation of real photonic corrections is based on full $2\to 5$ matrix elements as in the {\tt RacoonWW} approach.
Photon--(anti-)quark induced subprocesses, as depicted in \reffi{fig:real}, are also taken into account.

The relative electroweak corrections to the main production channel $\bar q q \to \mvev$ with respect to
its LO prediction is denoted by $\de_{\bar q q}$ in the discussion of numerical results.
As the $\ga q$- or $\bar q \ga$-induced subprocesses were often neglected 
in previous studies of EW corrections to related processes
due to the small size of the photon distribution
function, which in addition is plagued by large uncertainties, we study the impact of their contribution,
denoted by $\de_{q\ga}$, separately.

Finally, we summarize our predictions in the relative correction factor $\de_{\rm{EW}}$ given by
\bce
$\de_{\rm{EW}}=\de_{\bar q q}+\de_{q\ga}+\de_{\bar\Pb\Pb}+\de_{\ga\ga}\,$.
\ece 

\subsection{Building blocks of the Monte Carlo program}
All matrix elements required in the calculation are covered by inhouse routines.
The tree-level matrix elements
are calculated in the Weyl-van-der-Waerden spinor formalism leading to compact results.
The loop amplitudes are generated using {\tt FeynArts}~\cite{Kublbeck:1990xc} and inhouse {\tt Mathematica} routines
resulting in {\tt Fortran} code. This code is evaluated with the help of the loop library 
{\tt Collier} whose implementation relies on the results presented in \citeres{Passarino:1979jh,'tHooft:1979xw,Dittmaier:2003bc}.
The numerical integration is performed by an adapted version of the multi-channel phase-space generator
in the Monte Carlo program {\tt Coffer$\ga\ga$}~\cite{Bredenstein:2005zk}.

\section{Numerical results}
%
\btab
\begin{center}
\begin{tabular}{|c|c|cccc|}
\hline
&$\si_{\bar q q}^\mr{LO}$~[fb] &  $\de_{\bar q q}$~[$\%$]& $\de_{q \gamma}$~[$\%$] & $\de_{\gamma\gamma}$~[$\%$] & $\de_{\bar\Pb\Pb}$~[$\%$]
\\
\hline
\hline
LHC14     &412.5(1) & $-2.70(2)$ & $0.566(5)$   & $0.7215(4)$  & $1.685(1)$ 
\\
\hline
LHC8      &236.83(5) & $-2.76(1)$ & $0.470(3)$   & $0.8473(3)$  & $0.8943(3)$
\\
\hline
ATLAS cuts&163.84(4) & $-2.96(1)$ & $-0.264(5)$  & $1.0221(5)$  & $0.9519(4)$ 
\\
\hline
\end{tabular}
\caption{
\label{tab:xsecs}
Cross-section contributions to $\Pp\Pp\to \mvev$ at the LHC with centre-of-mass energies of $14\TeV$ (first line) and $8\TeV$ (second line), respectively. The third line shows the corresponding results for a collider energy of $8\TeV$ with the ATLAS setup. The numbers in brackets represent the numerical error on the last given digit. (Taken from \citere{Billoni:2013aba}.)  
}
\end{center}
\etab
%
%
The numerical results presented in the following are obtained using the parton distribution functions (PDFs) of the {\tt NNPDF2.3QED}~\cite{Ball:2013hta}~set
which provides also a photon distribution function. Furthermore, we recombine final-state leptons and nearly
collinear photons in order to obtain IR-safe observables. The recombination procedure and all relevant input
parameters are described in \citere{Billoni:2013aba}.

In \refta{tab:xsecs} we present the integrated LO cross section of the $\bar qq$-initiated process, $\si_{\bar q q}^{\rm LO}$, and the
respective relative correction factors for three different LHC scenarios.
While in the scenarios ``LHC14'' and ``LHC8'' only standard cuts ($p_{\rT,\ell}>20\GeV$ and $|y_\ell|<2.5$) 
on the charged final-state leptons are applied, the selection criterion in the scenario ``ATLAS cuts'' corresponds to
realistic acceptance cuts (see \citere{Billoni:2013aba}).
The cross section at LO accuracy is dominated by contributions initiated by light quarks, $\si_{\bar q q}^\mr{LO}$. Subprocesses with bottom quarks in the initial state yield an additional contribution of less than 2\% for the three setups under investigation. 
Only about 1\% of the full LO cross section stems 
from the photon-initiated contributions.
This result rectifies a posteriori the neglect of order $\mc{O}(\alpha)$  corrections to subprocesses of the type $\bar\Pb\Pb \to\mvev$ 
and $\gamma\gamma\to\mvev$.

Apparently, the sum of all considered corrections is very small as the small negative EW corrections to the quark-initiated processes are widely
compensated by positive corrections of the separately considered LO contributions. However, the EW corrections significantly distort 
distributions, since they are not uniformly distributed in phase space, but tend to increase at scales above the weak-boson mass. 
%
\bfi
\includegraphics[angle=0,scale=0.7,bb=60 330 370 770]{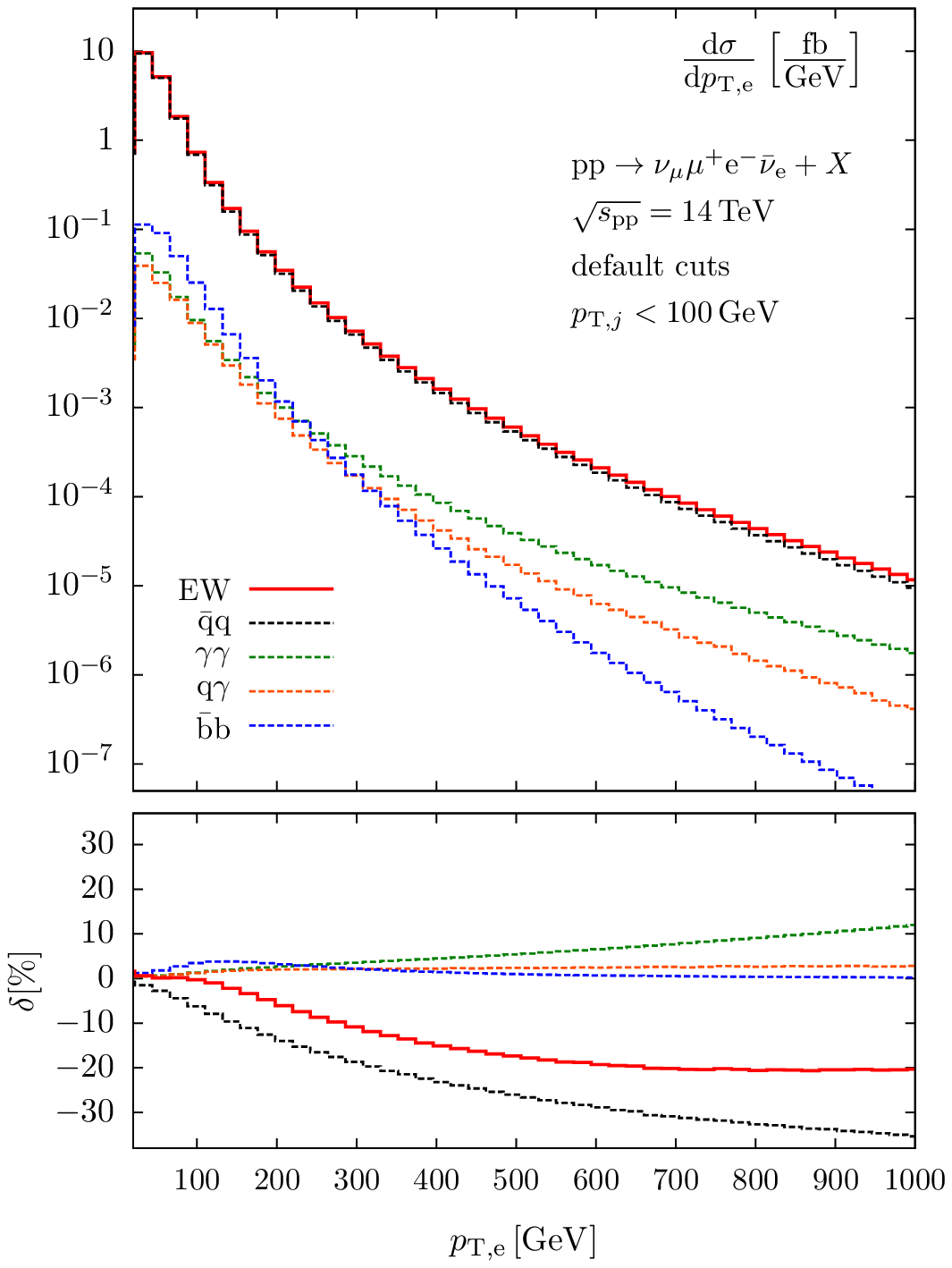}
\includegraphics[angle=0,scale=0.7,bb=60 330 370 770]{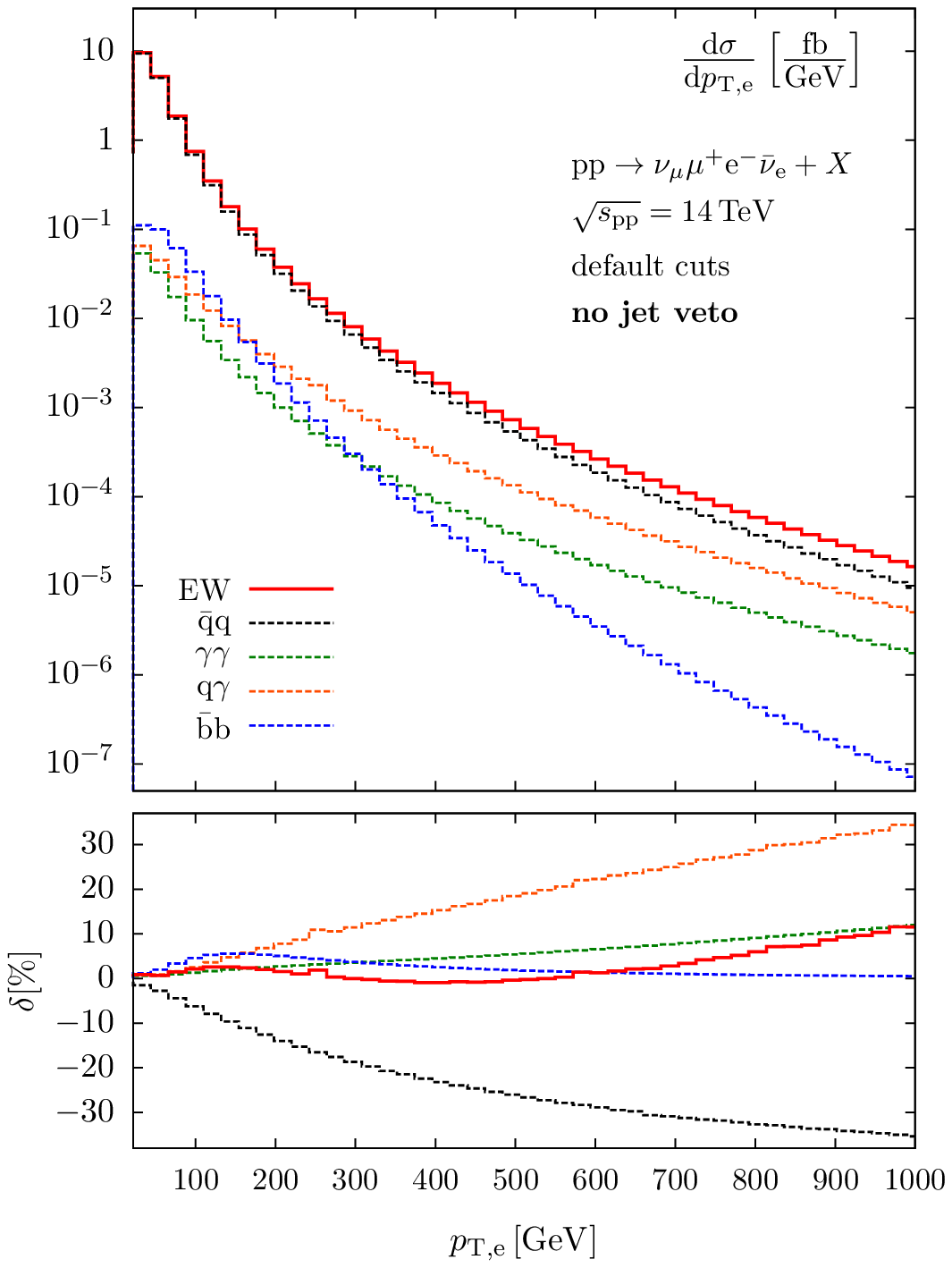}
\caption{Transverse-momentum distribution of the electron in $\Pp\Pp\to\mvev$ at NLO~EW accuracy (upper panels) for our default setup at the LHC14 with a jet veto of $100\GeV$ (left) and without a jet veto (right),  
together with the relative impact of individual contributions in each case (lower panels). (Taken from \citere{Billoni:2013aba}.) 
\label{fig:pte}
}
\efi
%

The transverse-momentum distribution of the electron receives large negative EW corrections at high transverse momentum
(black curve in \reffi{fig:pte}).
We note that a much stronger increase of the quark--photon induced contributions at large $p_{\rT,\Pe}$ is found in the absence of a jet veto. This feature is illustrated by \reffi{fig:pte}~(right) 
that shows the quark--photon contribution as a function of $p_{\rT,\Pe}$ for the same setup as in \reffi{fig:pte}~(left), apart from the jet veto $p_{\rT,j}<100\GeV$ . Normalization and shape of all contributions that do not contain a QCD 
parton in the final state, and therefore cannot give rise to a jet, are identical to the case where a jet veto is imposed. However, a significant increase in the relative size of the quark--photon contributions can be observed in the tail of the transverse-momentum distribution, giving rise to a relative contribution $\de_{q\gamma}$ of about
30\% for $p_{\rT,\Pe}=0.9\TeV$. 
This effect is due to 
the special enhancement in the amplitudes with initial-state photons coupled to $\PW$ bosons in $t$ channels, amplified by some recoil
against a hard jet in the final state. 
Naturally, QCD radiative corrections are enhanced by similar recoil effects and increase dramatically in the investigated kinematic region.
We conclude that a pure NLO prediction is not adequate to describe the tail of the transverse-momentum distributions,
unless a jet veto is applied.

%
\bfi
\includegraphics[angle=0,scale=0.7,bb=60 330 370 770]{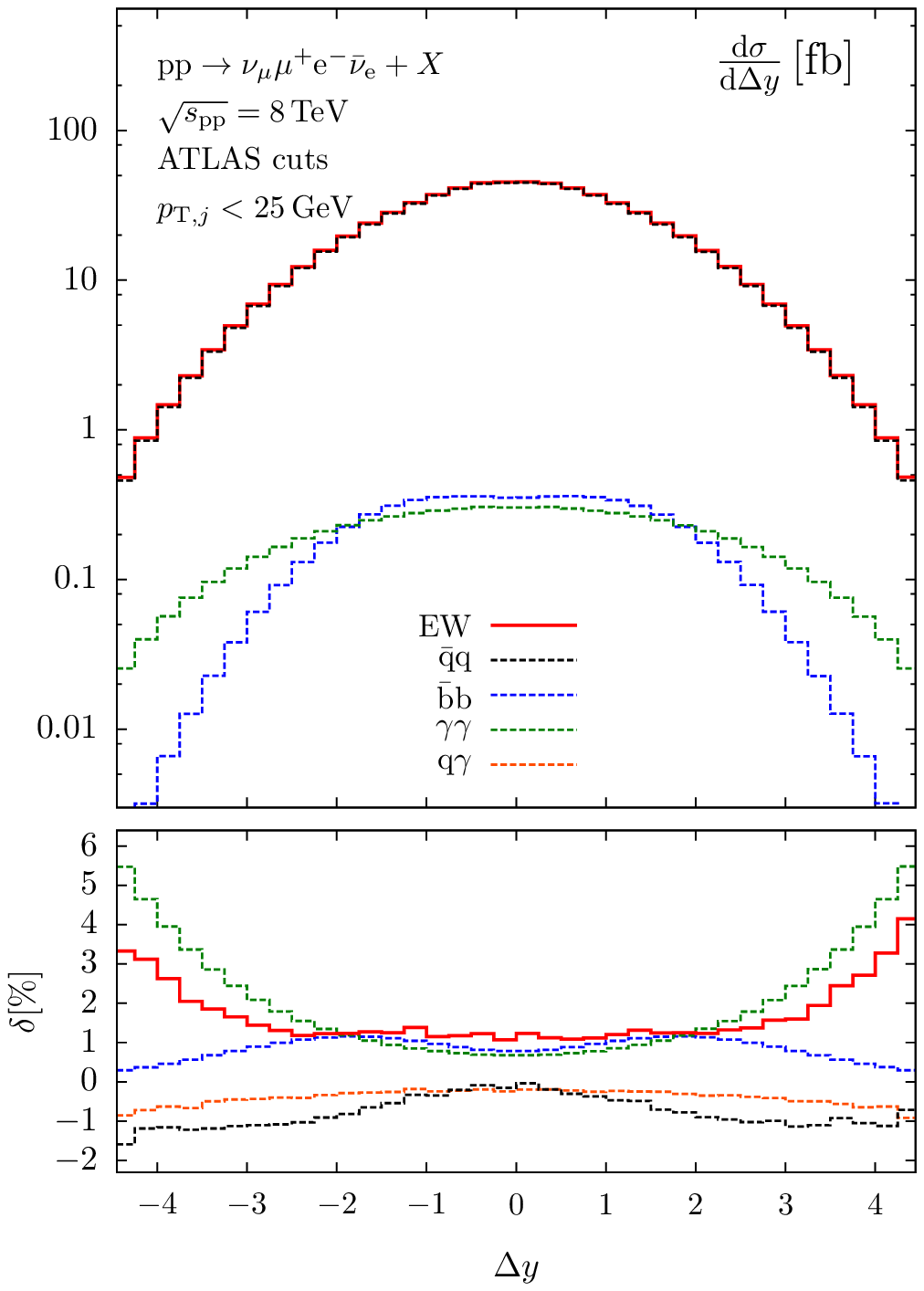}
\includegraphics[angle=0,scale=0.7,bb=60 330 370 770]{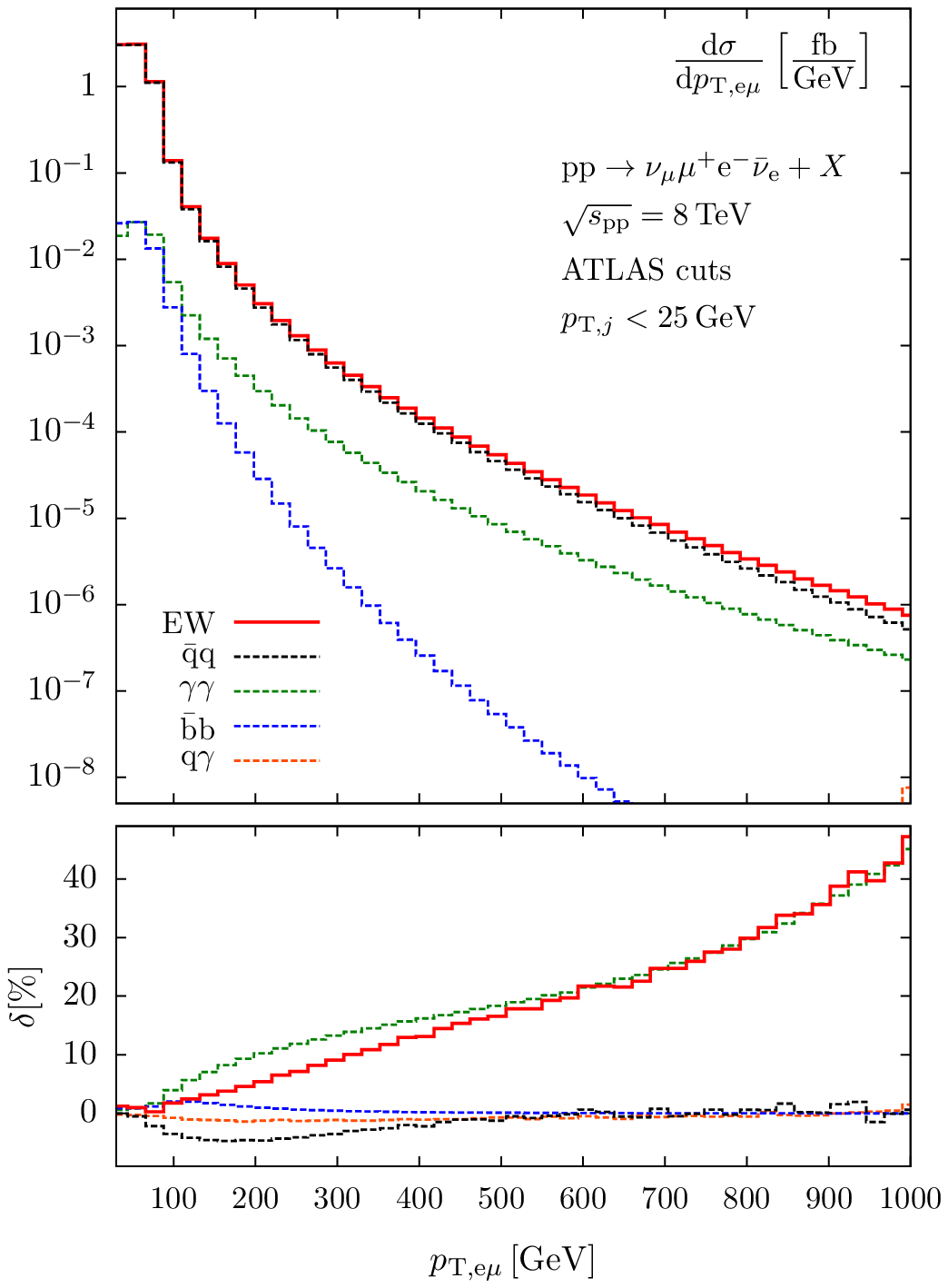}
\caption{Rapidity difference (left) and transverse-momentum distribution of the electron--muon system (right) 
in $\Pp\Pp\to\mvev$ at NLO~EW accuracy (upper panels) with realistic selection cuts at a collider energy of $8\TeV$,  
together with the relative impact of individual contributions in each case (lower panels). (Taken from \citere{Billoni:2013aba}.)
\label{fig:angular}
}
\efi
%
As an example for results on angular distributions in \reffi{fig:angular}~(left) we present the charged lepton rapidity
difference 
$\De y=y_\mu-y_\Pe$. The corrections slightly increase for
back-to-back configurations of the two charged leptons and the photon--photon induced contribution dominates the forward--backward 
emission of the charged leptons. Nevertheless, the sum over all contributions remains small, leading to total corrections of less than $5\%$. 

The NLO EW corrections to the transverse momentum of the charged lepton system (r.h.s of \reffi{fig:angular}) 
show a completely different behaviour. 
For this observable the negative EW corrections to the $\bar q q$-induced channels are compensated by the large impact 
of the process $\Pp\Pp\to\PWp\PWm\ga$ with real radiation of a hard photon, 
since a large photon recoil allows for
higher values of the transverse momentum of the W~pair which is effectively transferred to the decay leptons
due to the strong boost of the decaying W~bosons.
It turns out that in the current setup 
the photon--photon induced contribution dominates our predictions for the total correction at high $p_{\rm T,\Pe\mu}$.

\section{Conclusion}
We have presented results on next-to-leading-order electroweak corrections to the process $\Pp\Pp\to\mvev$ in the {\tt RacoonWW} approach where the virtual corrections are calculated in double-pole approximation while no approximation or simplification is made in the real-emission part. Individual contributions have been discussed in detail for total and differential cross sections.
The photon--photon and quark--photon induced contributions amount to $5-10\%$. The latter contribution
is suppressed by applying a jet veto that is necessary to avoid huge QCD corrections. The large negative electroweak corrections
to quark--antiquark scattering
at high scales are partially compensated by these two contributions.
Nevertheless, the total electroweak corrections can reach several tens of percent in kinematic regions of high momentum transfer where ``new physics'' might be expected, and therefore the inclusion of electroweak corrections in the analysis of experimental data will be mandatory in the future.

\end{document}